\documentclass[aps,prl,onecolumn,showpacs,preprintnumbers,superscriptaddress]{revtex4}
\usepackage{bm,amsmath,amssymb,latexsym,mathrsfs,enumerate}
\usepackage[mathcal]{euscript}
\usepackage[hypertex]{hyperref}
\usepackage{graphicx}

\begin{document}

\title{STATIONARY PRECESSION TOPOLOGICAL SOLITONS WITH NONZERO HOPF INVARIANT IN A UNIAXIAL FERROMAGNET}

\author{A. B. Borisov}
\email{Borisov@imp.uran.ru}
\affiliation{Institute of Metal Physics, Ural Division, Russian Academy of Sciences, Yekaterinburg 620041, Russia}

\author{F. N. Rybakov}
\email[Corresponding author. Electronic address: ]{F.N.Rybakov@gmail.com}
\affiliation{Institute of Metal Physics, Ural Division, Russian Academy of Sciences, Yekaterinburg 620041, Russia}

\date{30 June, 2008}
\begin{abstract}
Three-dimensional stationary precession solitons with nonzero Hopf indices are found numerically by solving
the Landau–Lifshitz equation. The structure and existence domain of the solitons are found.
\end{abstract}


\maketitle

Topological three-dimensional solitons attract considerable
interest in many fields of physics including
hydrodynamics, particle physics, cosmology, and condensed
matter physics. In models with the three-component
unit vector field ${\bf n}=(n_{1}, n_{2}, n_{3})$, where ${\bf n}^{2}=1$, localized structures exist if the field
$n$ asymptotically approaches the vector   ${\bf n}_{0}$ as  $|{\bf r}|\to\infty$. Such fields map
the $R^3\cup \{\infty \}$ space to the two-dimensional sphere $S^2$  êand are classified by the homotopy classes $\pi_3(S^2)=Z$ and characterized by the Hopf invariant $Í$~\cite{bib:BW} given by
the expression:
\begin{equation}
H=-\frac{1}{(8\pi)^2}\int{\bf F}\cdot{\bf A} d{\bf r},\label{eq:Hdefault}
\end{equation}
where $F_{i}=\epsilon_{ijk}{\bf n}\cdot(\bigtriangledown_j{\bf n}\times \bigtriangledown_k{\bf n})$ and $curl{\bf A}=2{\bf F}$. The invariant $H$
admits the simple geometric interpretation
as the linking number of two preimage closed curves
corresponding to an arbitrary pair of points on the
$S^2$ sphere.

\begin{figure}
\includegraphics[width=0.8\columnwidth, viewport=80 505 370 720, clip]{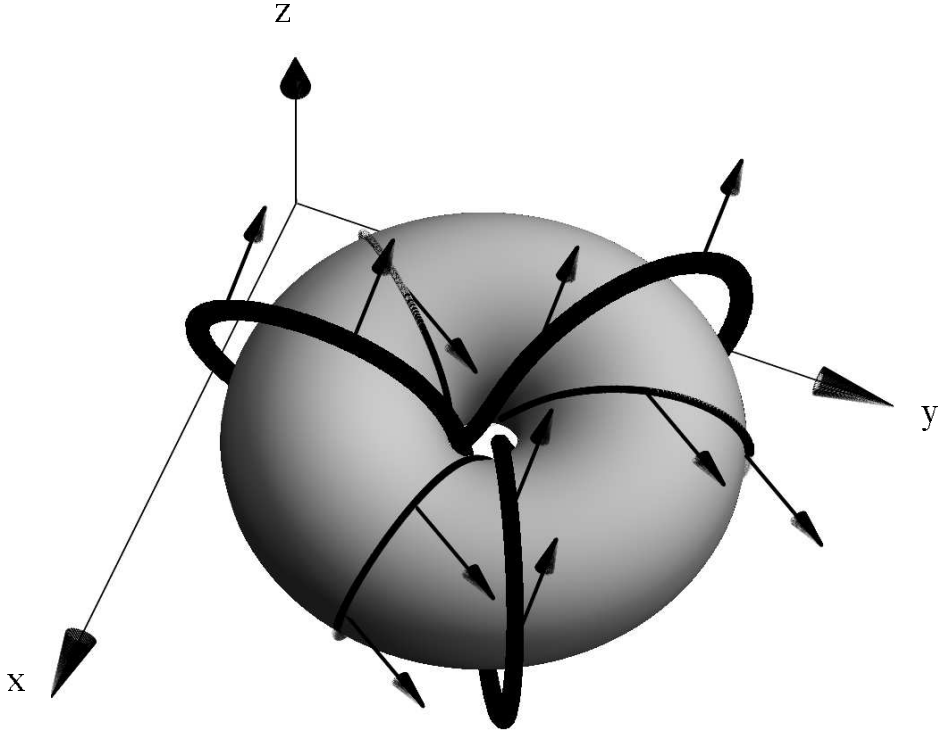}
\caption{Distribution of magnetization in a three-dimensional topological soliton with $H=3$. The thin and thick curves are
the preimages of the points $(\Theta_1, \Phi_1) = (0.5\pi, 1)$ and $(\Theta_2, \Phi_2) = (0.2\pi, 3)$ on the $S^2$ sphere, respectively.}
\label{F:fig1}
\end{figure}

Stable three-dimensional solitons with $H\neq0$ (the
so-called knotted solitons) were studied numerically in
the Faddeev-Niemi model~\cite{bib:LD,bib:LN,bib:LN1,bib:BS,bib:Glad}, i.e., the nonlinear
$\sigma$-model including terms with fourth-order derivatives.
Defects with a nonzero Hopf invariant have been discussed
in condensed matter physics since the pioneering
works by Volovik and Mineev for superfluid
 $^3He$~\cite{bib:VM} and by Dzyaloshinskii and Ivanov~\cite{bib:DI} for a uniaxial ferromagnet.
Simple reasoning based on the Derrick theorem~\cite{bib:Derrick} indicates the absence of nontrivial static threedimensional
solitons with finite energy in the aforementioned
media. However, dynamical structures of
this kind, which are stabilized by the precession of the
magnetization, can exist~\cite{bib:KBK,bib:PT}. Studies of threedimensional
magnetic structures are of both academic
and engineering interest in view of the development of new memory elements based on topological solitons in
uniaxial ferromagnets~\cite{bib:OBS}.

Three-dimensional magnetic structures have been
poorly studied until recently. A few original theoretical
and experimental works on magnetic structures in
superfluid $^3He$ were published in recent years~\cite{bib:Volovik1,bib:Volovik2,bib:Volovik3,bib:Volovik4}.
Spiral and cnoidal hedgehogs with ñ $H=0$ were analytically
described in ~\cite{bib:Borisov} in the Heisenberg model for an
isotropic ferromagnet. Constant-velocity precession solitons with
$H=1$ were numerically found in the same
model~\cite{bib:Cooper}. Both stationary (magnon drops) ~\cite{bib:IvKos1} and
uniformly moving~\cite{bib:Sut2001} precession radially symmetric
nontopological solitons in a uniaxial ferromagnet were
analyzed. In addition, the moving topological soliton
was recently analyzed~\cite{bib:Sut2007}, but stable configurations
with $H\neq0$ were not found.

In this work, we find dynamical structures, namely,
stationary precession three-dimensional solitons with
nonzero Hopf indices, in a uniaxial ferromagnet. Hereafter,
the existence domain of such solitons is determined
and their fine structure and main features are
studied.

The dynamics of the magnetization vector is
described by the Landau-Lifshitz equation. In the case
of negligible relaxation, this equation has the form
\begin{equation}
\frac {\partial {\bf M}} {\partial t} = - \frac {2 \mu _ 0} {\hbar}\left[\frac {\delta { E}} {\delta {\bf M}}\times {\bf M}\right],\label{eq:LLeq}
\end{equation}
where the ferromagnet energy  $E=E_{exch.}+E_{anis.}$ is the
sum of the exchange energy $E_{exch.}$ and the magnetic anisotropy
energy  $E_{anis.}$ with the parameter $\beta$:
\begin{align}
E_{exch.}&=\frac{\alpha }{2}\int \left(\frac {\partial {\bf M}} {\partial {r _ i}}\right)^2 d{\bf r},\label{eq:ExchEnergy}\\
E_{anis.}&=\frac{\beta }{2}\int \left({M_ x}^2+ {M_ y}^2\right) d{\bf r}.\label{eq:AnizEnergy}
\end{align}
The energy of the magnetic-dipole interaction is negligible
for large $\beta$ values typical for many magnetic media.

Using the parameterization
\begin{equation}
\begin{cases}
{\bf M}=M_{0} {\bf n},\\
{\bf n}=(sin \Theta cos\Phi, sin\Theta sin\Phi, cos\Theta),
\end{cases}\label{eq:Mparam}
\end{equation}
we seek solutions describing stationary precession solitons of the form
\begin{equation}
\begin{cases}
\Phi=\omega t + Q \varphi + \phi(r,z),\\
\Theta=\theta(r,z),
\end{cases}\label{eq:Qvortex}
\end{equation}
where $\varphi$ is the polar angle of the cylindrical coordinate
system $(r,\varphi,z)$ and $\phi$ and $\theta$ are the odd and even functions
of $z$, respectively, and the $z$ axis is the symmetry
axis. It is known~\cite{bib:Tjon,bib:PT} that studying the dynamics of
such structures is equivalent to solving a variational
problem of minimizing the energy functional $E$ for a
fixed invariant of motion, $N$, equal to the total number
of spin deviations:
\begin{equation}
N=\frac {M_0} {2 \mu _ 0}\int \left(1- n_z\right) d{\bf r}=const.\label{eq:Nintegral}
\end{equation}

The precession frequency is determined by the formula
\begin{equation}
\omega=\frac {1} {\hbar}\frac {\partial E} {\partial N}.\label{eq:omega1}
\end{equation}
The Derrick method~\cite{bib:Derrick} applied to finite-energy field
configurations yields the convenient formula for the
precession frequency:
\begin{equation}
\omega=\frac {1} {\hbar}\frac {E_{exch.} +3  E_{anis.}} {3 N}\label{eq:omega2}
\end{equation}

In view of Eqs.~(\ref{eq:Qvortex}), the energy functional takes the form
\begin{equation}
E=\int_{-\infty}^{\infty} \int_{0}^{\infty} w_E d{r}d{z},\label{eq:4}
\end{equation}
Here,
\begin{equation}
w_E=  \alpha \pi M_0^2 \left[ r { \left( \frac {\partial {\bf n^*}} {\partial r} \right)}^2 + r {\left(\frac {\partial {\bf n^*}} {\partial z} \right)}^2 +\left( \frac {Q^2} {r}+{\frac \beta \alpha} r \right) ({n_x^*}^2+{n_y^*}^2) \right],\label{eq:wE}
\end{equation}
where
\begin{equation}
{\bf n^*}=(sin \theta cos \phi, sin \theta sin \phi, cos \theta ) .\label{eq:nParam}
\end{equation}

\begin{figure}
\includegraphics[width=0.6\columnwidth, viewport=80 460 370 720, clip]{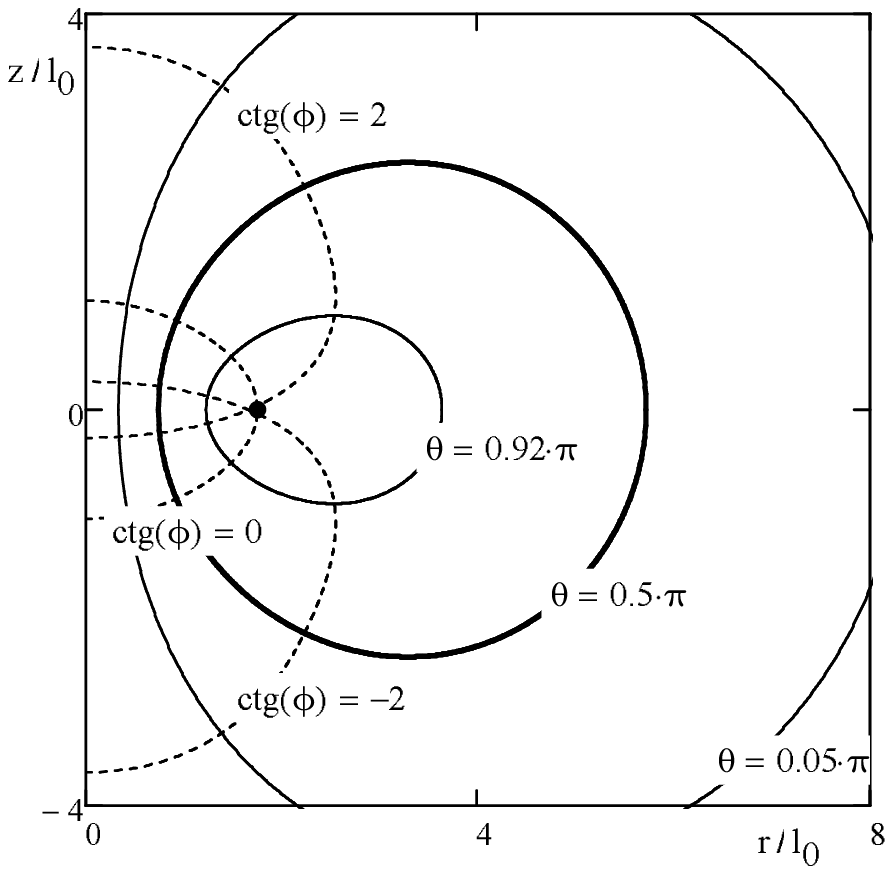}
\caption{Contours of the angles parameterizing the unit vector
${\bf n^*}$ for $H = 3$ and $\omega/\omega_0=0.5$. The solid and dashed curves
correspond to $\theta=const$ and $ctg(\phi)=const$, respectively.}
\label{F:fig2}
\includegraphics[width=0.7\columnwidth, viewport=15 610 310 760, clip]{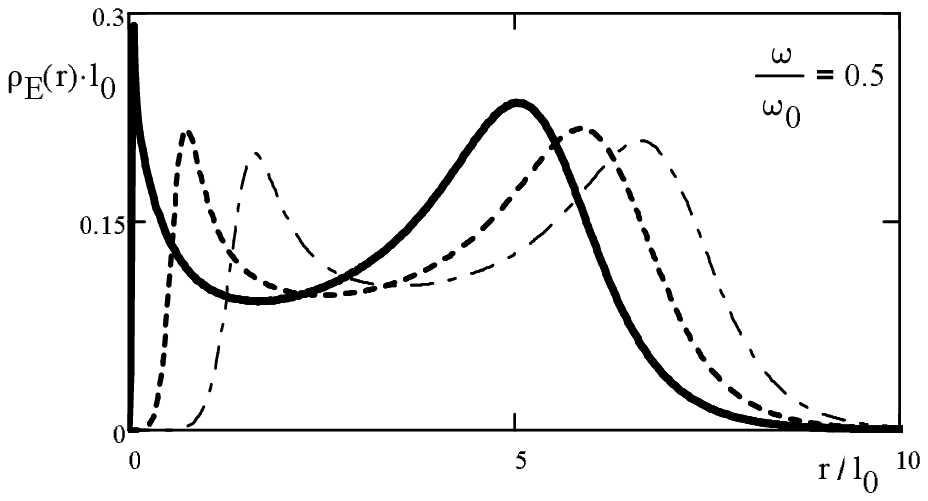}
\caption{Distance dependence of the normalized energy density
for solitons with the same precession frequency for various $H$ values. The solid, dashed, and dash–dotted curves correspond to $H = 2$, $3$, and $4$, respectively.}
\label{F:fig3}
\end{figure}
\begin{figure}
\includegraphics[width=0.6\columnwidth, viewport=20 310 300 765, clip]{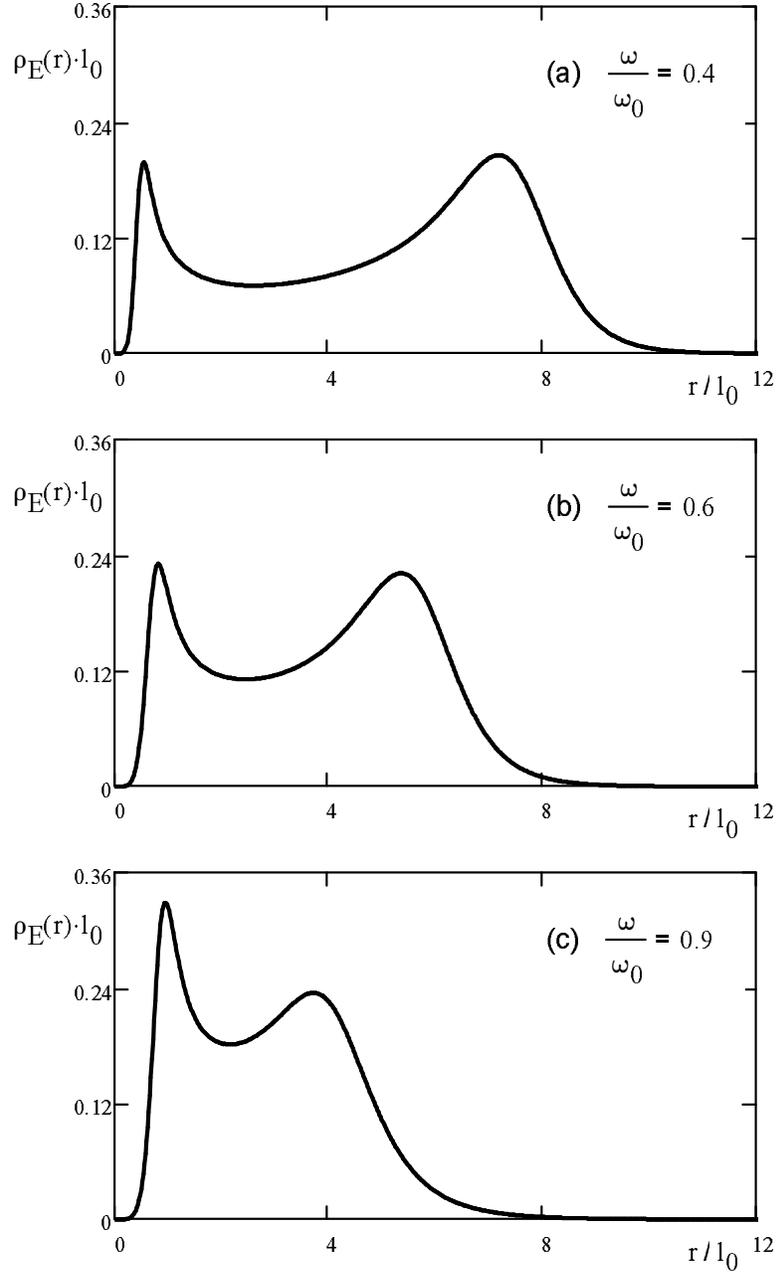}
\caption{Distance dependence of the normalized energy density
of a soliton for several values of the precession frequency and $H = 3$.}
\label{F:fig4}
\end{figure}

\begin{figure}
\includegraphics[width=0.6\columnwidth, viewport=20 495 300 690, clip]{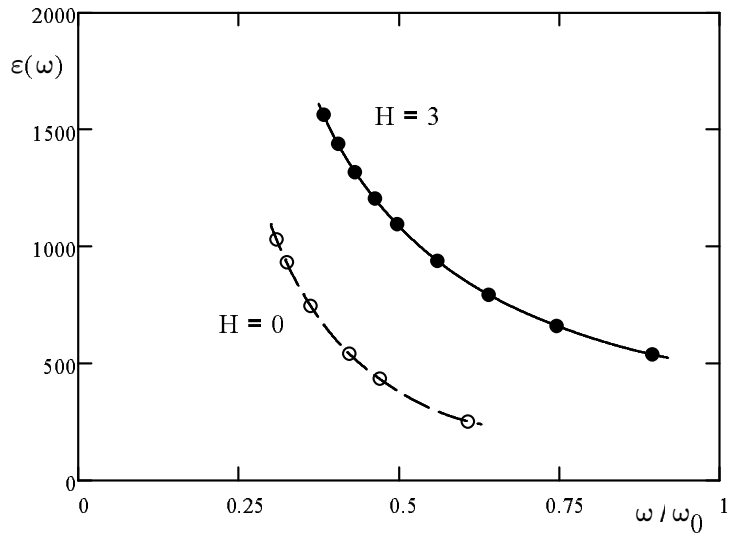}
\caption{Energy versus the precession frequency for nontopological
($H = 0$) and topological ($H = 3$) solitons. The points show the numerical results.}
\label{F:fig5}
\end{figure}

The formula describing the Hopf index simplifies
drastically for fields specified by Eqs.~(\ref{eq:Qvortex})~\cite{bib:KUR,bib:Glad}
\begin{equation}
H=Q T,\label{eq:HopfIndex}
\end{equation}
where the integer $T$ is given by
\begin{equation}
T=\frac {1} {4 \pi}\int_{-\infty}^{\infty} {\int_{0}^{\infty} {{\bf n^*}\cdot\left[\frac {\partial {\bf n^*}} {\partial r}\times \frac {\partial {\bf n^*}} {\partial z} \right]} d{r}}d{z}.\label{eq:TNumber}
\end{equation}

To solve the variational problem numerically by the
finite difference method, the energy functional was
replaced by a function of the projections of the vector ${\bf n^*}$
at the numerical-grid nodes. Since the field configuration
for negative $z$ values is determined by the symmetry
of the problem, the discretization domain was the
unit square $0 \leq r \leq 1,0 \leq z \leq 1$.

The expected domain of soliton localization should
be about the typical scale
\begin{equation}
l_0=\sqrt{\frac{\alpha}{\beta}}.\label{eq:L0}
\end{equation}

The parameters $\alpha$ and $\beta$ were chosen to ensure that $1/l_0\sim\sqrt{D}$, where $D$ is the number of grid nodes in one
direction. The key point of this method is to ensure that
the soliton is entirely contained in the simulation
domain, but is sufficiently large to be adequately
described by the discrete grid. In addition, to improve
the performance, the numerical grid was nonuniform
with the density of nodes increasing toward the coordinate
origin. This means that the inhomogeneity was
assumed to be localized near the center, while the field
quickly reaches the asymptotic uniform state at infinity.
The vector ${\bf n_{0}}=(0,0,1)$ at the boundaries $z=1$ or $r=1$
of the domain corresponds to the ground state.

The energy with an additive penalty function was
minimized by the conjugate gradient method. A quadratic
penalty function imposes a constraint on the
norm of the vector ${\bf n^*}$ which is unity at any point. Integral~(\ref{eq:Nintegral}) is discretized to a linear form, and the imposed
constraint is taken into account by applying a linear
operator~\cite{bib:PD} in the calculation of the direction toward
the minimal objective function for each iteration.

A feature of our numerical analysis was the choice
of the initial state, i.e., the trial field configuration from
which the minimization begins. The vector corresponded
to the ground state in the entire domain of simulation
excluding a small quadrant near the coordinate
origin. The vectors inside the quadrant lied in the
orthogonal plane, and their directions varied randomly
from point to point. Thus, the initial configuration is to
a certain extent chaotic and does not even obey the criteria
for the discretization of any smooth function.

The result was directly tested in addition to the
observed approach of the energy to its asymptotic value
after many iterations. The frequency $\omega$ was calculated
using Eq.~(\ref{eq:omega2}). The second derivatives and the discrepancy
of Landau-Lifshitz equation~(\ref{eq:LLeq}) were numerically
evaluated at each internal point of the simulation
domain.

For the solitons presented below, the numerical calculations
based on Eq.~(\ref{eq:TNumber}) yielded $T = 1$. In this case,
according to Eq.~(\ref{eq:HopfIndex}), the Hopf index $H = Q$. Topological
solitons of other types will be described elsewhere.
The results of our present study are discussed below.

Figure \ref{F:fig1} shows a typical configuration of the field $\bf{n}$
for $H = 3$. Here, the coordinate axes are shifted for clarity
and illustrate only the spatial orientation of the soliton.
It is seen that the linking number of the preimages
of two points of the $S^2$ sphere is equal to three. The toroidal
surface corresponds to $\theta=\pi/2$. Note that ${\bf n}={\bf n}_0$ on
the symmetry axis $r = 0$ and ${\bf n}\rightarrow{\bf n}_0$ for $r^2+z^2\rightarrow\infty$.

The contours of the angles parameterizing the vector
${\bf n^*}$ are plotted in Fig. \ref{F:fig2}. The curve $\theta=\pi/2$ is a section
of an axisymmetric surface of kind 1 (toroidal surface)
in which a large fraction of the soliton energy is
contained. The common point of the lines $\phi=const$ is
the south pole of the sphere $S^2$, at which $\theta=\pi$.

Analysis showed that it is most difficult to simulate
solitons with small H values. A series of efficacious
numerical results were obtained for $H = 3$ and $4$ using a
600x400 grid. However, in the case where $H = 2$, a
plausible result was obtained only for a 1600x800
grid. The cause is an interesting feature of the structure
of the studied class of objects. This feature becomes
apparent in numerical calculations of the normalized
energy density
\begin{equation}
\rho_E=\frac{1}{E}\int_{-\infty}^{\infty} w_E d{z}.\label{eq:EnergyDensity}
\end{equation}
The dependences of $\rho_E$ on $r$ for various values of $H$ are
plotted in Fig. \ref{F:fig3}. The maximum $\rho_E$ values are reached in the vicinity of $\theta=\pi/2$.
With a decrease in $H$, the localization
domain corresponding to the first maximum of
$\rho_E(r)$ becomes narrower and the maximum becomes
steeper. The soliton structure becomes narrower, so that
a finer grid with more nodes should be used.

Three-dimensional solitons with $H = 0$ exist for $\omega\leq0.915\omega_0$ ~\cite{bib:IvKos1}, where $\omega_0$ is the ferromagnetic-resonance frequency:
\begin{equation}
\omega_0=\frac {2 \mu_0 M_0 \beta} {\hbar}\label{eq:omega0}
\end{equation}
An increase in the precession frequency results in the
compression of the soliton (Fig.\ref{F:fig4}) implying the possible
existence of certain threshold ratio $\omega/\omega_0$ similar to
the nontopological soliton~\cite{bib:IvKos1}.

Figure \ref{F:fig5} shows the dimensionless energy $\varepsilon={E}/(\alpha l_0)$
as a function of $\omega/\omega_0$. It is seen that the soliton energy decreases
with an increase in the precession frequency.
A similar dependence for a magnon drop, calculated by
the aforementioned procedure, is shown in this figure
for a comparison.

The discovered solitary structures with typical sizes
of a few to a few tens of $l_0$ are much smaller than cylindrical
magnetic domains and, therefore, can potentially
be used for information recording and storage if mechanisms
for their generation and control will be developed
in the same way as it is done now for magnetic
vortices~\cite{bib:OBS}. This would provide the possibility of
recording information in three-dimensional samples.


\begin{thebibliography}{99}
$ $
\bibitem{bib:BW}
R. Bott and L.\,W. Too, Differential Forms in Algebraic Topology, 1982, New York.\\

\bibitem{bib:LD}
L.\,D. Faddeev, IAS Princeton, IAS-Report No.75-QS70, (1975).\\

\bibitem{bib:LN}
    \bibinfo{author}{L.~D. Faddeev},
    \bibinfo{author}{A.~J. Niemi},
    \bibinfo{journal}{Nature} \textbf{\bibinfo{volume}{387}},
    \bibinfo{pages}{58} (\bibinfo{year}{1997}).\\
    \url{http://arxiv.org/abs/hep-th/9610193}.\\

\bibitem{bib:LN1}
    \bibinfo{author}{L.~D. Faddeev},
    \bibinfo{author}{A.~J. Niemi},
    \bibinfo{journal}{hep-th/9705176},
    \bibinfo{pages}{} (\bibinfo{year}{1997}).\\
    \url{http://arxiv.org/abs/hep-th/9705176}.\\

\bibitem{bib:BS}
    \bibinfo{author}{R.~A. Battye},
    \bibinfo{author}{P.~M. Sutcliffe},
    \bibinfo{journal}{Phys. Rev. Lett.} \textbf{\bibinfo{volume}{81}},
    \bibinfo{pages}{4798} (\bibinfo{year}{1998}).\\
    \url{http://prola.aps.org/abstract/PRL/v81/i22/p4798_1},\\
    \url{http://arxiv.org/abs/hep-th/9808129}.\\

\bibitem{bib:Glad}
    \bibinfo{author}{J. Gladikowski},
    \bibinfo{author}{M. Hellmund},
    \bibinfo{journal}{Phys. Rev. D} \textbf{\bibinfo{volume}{56}},
    \bibinfo{pages}{5194} (\bibinfo{year}{1997}).\\
    \url{http://prola.aps.org/abstract/PRD/v56/i8/p5194_1},\\
    \url{http://arxiv.org/abs/hep-th/9609035}.\\

\bibitem{bib:VM}
    \bibinfo{author}{G.~E. Volovik},
    \bibinfo{author}{V.~P. Mineev},
    \bibinfo{journal}{Sov. Phys. JETP} \textbf{\bibinfo{volume}{46}},
    \bibinfo{pages}{401} (\bibinfo{year}{1977}).\\

\bibitem{bib:DI}
    \bibinfo{author}{I.~E. Dzyloshinskii},
    \bibinfo{author}{B.~A. Ivanov},
    \bibinfo{journal}{JETP Lett.} \textbf{\bibinfo{volume}{29}},
    \bibinfo{pages}{540} (\bibinfo{year}{1979}).\\
    \url{http://jetpletters.ac.ru/ps/1455/article_22156.shtml}\\


\bibitem{bib:Derrick}
    \bibinfo{author}{G.~M. Derrick},
    \bibinfo{journal}{J. Math. Phys} \textbf{\bibinfo{volume}{5}},
    \bibinfo{pages}{1252} (\bibinfo{year}{1964}).\\
    \url{http://link.aip.org/link/?JMAPAQ/5/1252/1}.\\

\bibitem{bib:KBK}
    \bibinfo{author}{A.~M. Kosevich},
    \bibinfo{author}{B.~A. Ivanov},
    \bibinfo{author}{A.~S. Kovalev},
    \bibinfo{journal}{Phys. Rep.} \textbf{\bibinfo{volume}{194}},
    \bibinfo{pages}{117} (\bibinfo{year}{1990}).\\
    \url{http://dx.doi.org/10.1016/0370-1573(90)90130-T}.\\

\bibitem{bib:PT}
    \bibinfo{author}{N. Papanicolaou},
    \bibinfo{author}{T.~N. Tomaras},
    \bibinfo{journal}{Nucl. Phys. B} \textbf{\bibinfo{volume}{360}},
    \bibinfo{pages}{425} (\bibinfo{year}{1991}).\\
    \url{http://dx.doi.org/10.1016/0550-3213(91)90410-Y}.\\


\bibitem{bib:OBS}
    \bibinfo{author}{T. Okuno},
    \bibinfo{author}{K. Mibu},
    \bibinfo{author}{T. Shinjo},
    \bibinfo{journal}{J. Appl. Phys} \textbf{\bibinfo{volume}{95}},
    \bibinfo{pages}{3612} (\bibinfo{year}{2004}).\\
    \url{http://link.aip.org/link/?JAPIAU/95/3612/1}.\\

\bibitem{bib:Volovik1}
    \bibinfo{author}{G.~E. Volovik},
    \bibinfo{journal}{cond-mat/0701180},
    \bibinfo{pages}{} (\bibinfo{year}{2007}).\\
    \url{http://arxiv.org/abs/cond-mat/0701180}.\\

\bibitem{bib:Volovik2}
    \bibinfo{author}{Yu.~M. Bunkov},
    \bibinfo{author}{G.~E. Volovik},
    \bibinfo{journal}{Phys. Rev. Lett.} \textbf{\bibinfo{volume}{98}},
    \bibinfo{pages}{265302} (\bibinfo{year}{2007}).\\
    \url{http://link.aps.org/abstract/PRL/v98/e265302},\\
    \url{http://arxiv.org/abs/cond-mat/0703183}.\\

\bibitem{bib:Volovik3}
    \bibinfo{author}{Yu.~M. Bunkov},
    \bibinfo{author}{G.~E. Volovik},
    \bibinfo{journal}{J. Low Temp. Phys} \textbf{\bibinfo{volume}{150}},
    \bibinfo{pages}{135} (\bibinfo{year}{2008}).\\
    \url{http://www.springerlink.com/content/a55n5722617l68k8},\\
    \url{http://arxiv.org/abs/0708.0663}.\\

\bibitem{bib:Volovik4}
    \bibinfo{author}{Yu.~M. Bunkov},
    \bibinfo{author}{G.~E. Volovik},
    \bibinfo{journal}{Physica C} \textbf{\bibinfo{volume}{468}},
    \bibinfo{pages}{609} (\bibinfo{year}{2008}).\\
    \url{http://dx.doi.org/10.1016/j.physc.2007.11.026},\\
    \url{http://arxiv.org/abs/0710.3448}.\\

\bibitem{bib:Borisov}
    \bibinfo{author}{A.~B. Borisov},
    \bibinfo{journal}{ JETP Lett.} \textbf{\bibinfo{volume}{76}},
    \bibinfo{pages}{84} (\bibinfo{year}{2002}),\\
    \url{http://www.springerlink.com/content/g48t670u4906110h}.\\

\bibitem{bib:Cooper}
    \bibinfo{author}{N.~R. Cooper},
    \bibinfo{journal}{Phys. Rev. Lett.} \textbf{\bibinfo{volume}{82}},
    \bibinfo{pages}{1554} (\bibinfo{year}{1999}),\\
    \url{http://link.aps.org/abstract/PRL/v82/p1554},\\
    \url{http://arxiv.org/abs/cond-mat/9901037}.\\

\bibitem{bib:IvKos1}
    \bibinfo{author}{B.~A. Ivanov},
    \bibinfo{author}{A.~M. Kosevich},
    \bibinfo{journal}{JETP Lett.} \textbf{\bibinfo{volume}{24}},
    \bibinfo{pages}{495} (\bibinfo{year}{1976}).\\
    \url{http://jetpletters.ac.ru/ps/1816/article_27757.shtml}.\\

\bibitem{bib:Sut2001}
    \bibinfo{author}{T. Ioannidou},
    \bibinfo{author}{P.~M. Sutcliffe},
    \bibinfo{journal}{Physica D} \textbf{\bibinfo{volume}{150}},
    \bibinfo{pages}{118} (\bibinfo{year}{2001}).\\
    \url{http://dx.doi.org/10.1016/S0167-2789(00)00221-9},\\
    \url{http://arxiv.org/abs/cond-mat/0101129}.\\

\bibitem{bib:Sut2007}
    \bibinfo{author}{P.~M. Sutcliffe},
    \bibinfo{journal}{Phys. Rev. B} \textbf{\bibinfo{volume}{76}},
    \bibinfo{pages}{184439} (\bibinfo{year}{2007}).\\
    \url{http://link.aps.org/abstract/PRB/v76/e184439},\\
    \url{http://arxiv.org/abs/0707.1383}.\\

\bibitem{bib:KUR}
    \bibinfo{author}{A. Kundu},
    \bibinfo{author}{Y.~P. Rybakov},
    \bibinfo{journal}{ J. Phys. A} \textbf{\bibinfo{volume}{15}},
    \bibinfo{pages}{269} (\bibinfo{year}{1982}).\\
    \url{http://www.iop.org/EJ/abstract/0305-4470/15/1/035}.\\

\bibitem{bib:Tjon}
    \bibinfo{author}{J. Tjon},
    \bibinfo{author}{J. Wright},
    \bibinfo{journal}{Phys. Rev. B} \textbf{\bibinfo{volume}{15}},
    \bibinfo{pages}{3470} (\bibinfo{year}{1977}).\\
    \url{http://link.aps.org/abstract/PRB/v15/p3470}.\\

\bibitem{bib:PD}
B.\,N. Pshenichnyi and Yu.\,M. Danilin, Numerical Methods in Extremal Problems (Nauka, Moscow, 1975) [in Russian].

\end{thebibliography}
\end{document}